%% file: hep.tex
\documentclass[12pt]{article} 
\usepackage{epsfig}
\usepackage[hang,bf,small]{caption}
\usepackage{amsmath}
\DeclareGraphicsExtensions{.eps.gz,.eps,.ps,.ps.gz}
\parskip=\itemsep               %?
\date{\empty}

\include{mydef}

\def\lsim{\mathrel{\rlap{\lower4pt\hbox{\hskip1pt$\sim$}}
    \raise1pt\hbox{$<$}}}                % less than or approx. symbol
\def\gsim{\mathrel{\rlap{\lower4pt\hbox{\hskip1pt$\sim$}}
    \raise1pt\hbox{$>$}}}                % greater than or approx. symbol
\begin{document}

\title{
{\normalsize\rightline{DESY 99-067}\rightline{MPI-PhE/99-02}
\rightline{hep-ph/9906441}} 
\vskip 1cm 
      \bf QCD Instanton-induced Processes\\ 
      \bf in Deep-inelastic Scattering - \\
      \bf Search Strategies and Model Dependencies\thanks{Contribution to 
the Proceedings of the DESY Workshop 1998/99 on Monte Carlo Generators for 
HERA Physics.}
       \vspace{11mm}}
\author{T. Carli$^a$, J. Gerigk$^a$, A. Ringwald$^b$ and F. Schrempp$^b$
        \\[4mm] 
$^a$ Max-Planck-Institut f{\"u}r Physik, M{\"u}nchen, Germany\\
$^b$ Deutsches Elektronen-Synchrotron DESY, Hamburg, Germany}
\begin{titlepage} 
  \maketitle
\begin{abstract}
We investigate possible search strategies for QCD-instanton induced
processes at HERA in the deep-inelastic scattering (DIS) regime. Our
study is based 
on the Monte Carlo generator QCDINS for instanton-induced events and
the standard generators for normal DIS events. It appears possible to isolate
an instanton enriched data sample via an optimized  multi-dimensional
cut scenario for a set of six most instanton-sensitive DIS observables.
As a further central point, we investigate the stability of our 
results with respect to a variation of the (hadronization) models
available for the simulation of both normal DIS and instanton-induced
events. Dependencies on the variation of certain inherent
parameters are also studied. Within the ``bandwidth'' of variations
considered, we find that the normal DIS background is typically 
much more sensitive to model variations than the I-induced signal.
\end{abstract} 
\thispagestyle{empty}
\end{titlepage}
\newpage \setcounter{page}{2}

%
%%%%%%%%%%%%%%%%%%%%%%%%%%%%%%%%%%%%%%%%%%%%%%%%%%%%%%%%%%%%%%%%%%%%%%%%%%
\section{Introduction}

Instantons (I) have been known to theoretical physics since the mid 
seventies~\cite{Polyakov,tHoofta,tHooftb,tHooftc}. They represent
non-perturbative
fluctuations of the gauge fields in non-Abelian theories like
QCD and weak interactions, associated with tunneling transitions
between degenerate ground states (vacua) of different topology.

Instanton transitions induce processes~\cite{tHoofta,tHooftb,tHooftc}  
violating the (classical) conservation of certain fermionic quantum
numbers, namely chirality in QCD and the sum of baryon and lepton
number in weak interactions. These processes are forbidden in ordinary
perturbation theory, but have to exist in general due to the famous
ABJ chiral anomaly \cite{adler,belljarckiw,bardeen}. 

The DIS regime at HERA offers a unique possible discovery window for events
induced by QCD-instantons through their characteristic final state signature 
\cite{RS1,qcdins1,RS3,RS6,jgerigk} and a sizable rate, 
calculable\footnote{For an exploratory calculation of the I-contribution
to the gluon structure function see Ref.~\cite{balitsky}.}
within I-perturbation theory~\cite{moch,RS4,moch2}.
An experimental detection of these processes would correspond to a
novel, non-perturbative manifestation of non-Abelian gauge theories
and would clearly be of basic significance.

This report is organized as follows\footnote{An extended version of this
paper will appear elsewhere~\cite{qcdins2}.}:

We start off in section \ref{qcdins} by introducing the dominant
I-induced DIS process along with the relevant kinematics. Then we
proceed by summarizing the essential ingredients and the basic structure of the
Monte Carlo generator QCDINS for instanton-induced DIS
events~\cite{qcdins1,qcdins3},  on which the
present study is based. In section~\ref{signature}, 
we recall the theoretical results for the I-induced cross section at HERA and
summarize the corresponding characteristic topology of the
hadronic final state.
Section~\ref{recon} is devoted to a study of the possibility to
reconstruct the Bjorken variables of the I-subprocess.
Sections~\ref{search} and \ref{depend} contain the central results of
this investigation.  
In Section~\ref{search}, we report on a possible search strategy for 
these
processes. The goal is to isolate  an instanton enriched data sample
via an optimized  multi-dimensional cut scenario for a set of most
I-sensitive DIS observables.
Finally, in section~\ref{depend}, we investigate the stability of our 
results with respect to a variation of the (hadronization) models
available for the simulation of both normal DIS and instanton-induced
events. Dependencies on the variation of certain inherent
parameters are also considered.

All the studies presented in this report are performed in the hadronic center
of mass frame\footnote{This frame is defined by
$\vec{q}+\vec{P}=\vec{0}$. Therefore, the photon and the incoming
proton collide head-on.} (hCMS), which is a suitable frame of reference
in view of a good distinction between
I-induced and normal DIS events (c.\,f. \cite{jgerigk}).
The results that we obtain are based on a study of the hadronic final
state, with typical acceptance cuts of a HERA detector being applied 
in the laboratory frame
($15^\circ < \Theta_{hadron} <155^\circ$ and $p_T(hadron) > 0.15$ GeV,
for charged particles,
$\Theta_{hadron} >4^\circ$, for all stable particles, and
$0.25<p_T^2(K^0) <4.5~ \mbox{GeV}^2$, for neutral kaons)\footnote{
In the laboratory frame the incoming proton points in the positive 
z-direction, while in the hCMS the proton points in the negative 
z-direction.}. 
%$15^\circ < \Theta_{hadron} <155^\circ$ and $p_T(hadron) > 0.15$ GeV
%for a more realistic representation of the acceptance of the central
%tracking chamber of the H1 detector, and $0.25<p_T^2(K^0) <4.5~ \mbox{GeV}^2$,
%necessary for the detection of neutral kaons (via secondary vertices).
%
%%%%%%%%%%%%%%%%%%%%%%%%%%%%%%%%%%%%%%%%%%%%%%%%%%%%%%%%%%%%%%%%%%%%%%%%%%%%

\section{The Monte Carlo Generator QCDINS \label{qcdins}}

\begin{figure}[h]
\vspace{0.5cm}
%   \centering
%\hspace{1.0cm}
\begin{tabular}{ll}
\mbox{
 \epsfig{file=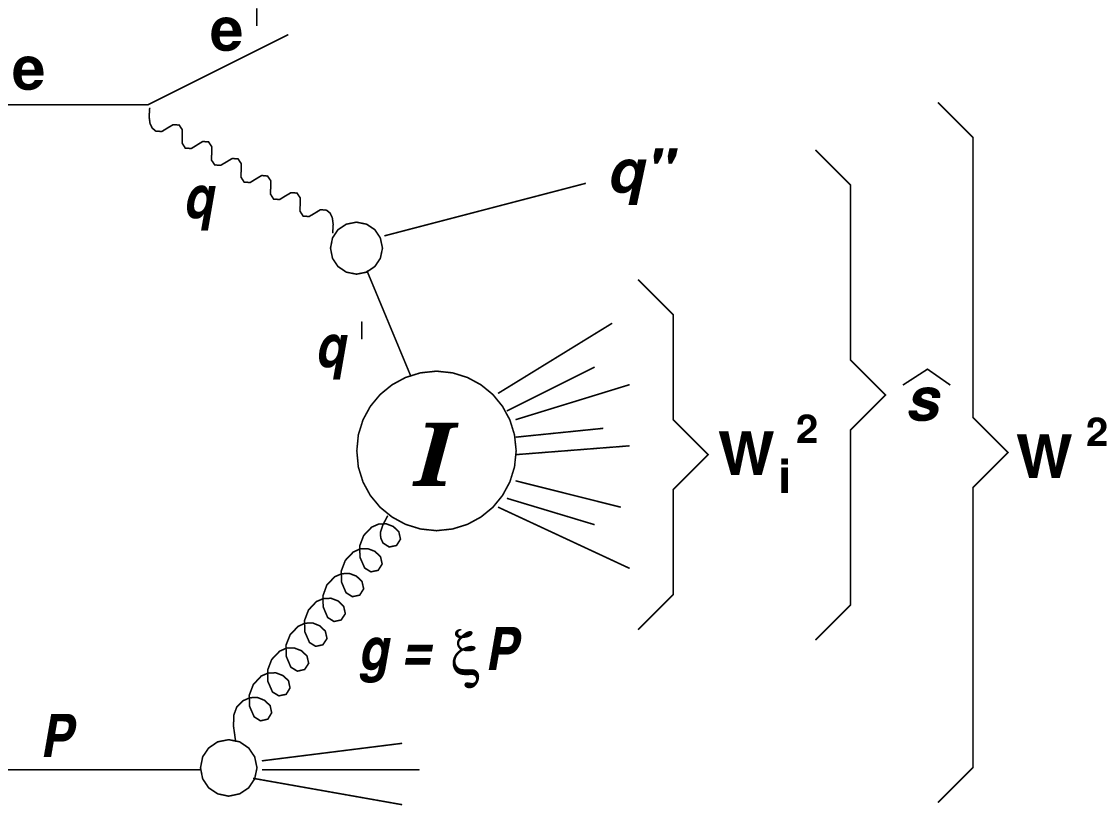,width=7cm,%
 bbllx=80pt,bblly=355pt,bburx=410pt,bbury=600,clip=}}
&
%\mbox{\vspace{-7.cm}
\begin{tabular}{l}
 \vspace{-5.5cm} \\
 DIS variables: \\
   $S=(e+P)^2$\\
   $\Qsq = - q^2 = -(e-e')^2$ \\
   $ x_{\rm Bj} = \Qsq / \; (2 P \cdot q) $ \\
   $ y_{\rm Bj} = \Qsq / \; (S\, x_{\rm Bj}) $\\
   $ W^2 =(q+P)^2 = \Qsq (1 - x_{\rm Bj})/x_{\rm Bj}$ \\
   $ \hat{s} = (q+g)^2$ \\
   $ \xi = x_{\rm Bj} \;(1+\hat{s}/Q^2)$ \\ \\
 Variables of I-subprocess: \\
 $\qprimesq=- {q'}^2 = - (q-q'')^2 $ \\
 $x'= \qprimesq / \;(2 \; g \cdot q' ) $ \\
 $W_i^2= (q'+g)^2 = \qprimesq ( 1 - \xprime )/ \xprime $\\
% $0 \leq x_{\rm Bj} \leq x_{\rm Bj}/\xi \leq x' \leq 1$
%}
\end{tabular}
%}
\end{tabular}
\vspace{0.3cm}
   \caption[Kinematic variables of I-induced process
     in deep-inelastic scattering]
     {Kinematic variables of the dominant I-induced process
     in deep-inelastic scattering.} 
%      The virtual photon $\gamma$ 
%     (4-momentum $q=e-e'$), emitted by the incoming electron,
%     fuses with a gluon (4-momentum $g$) of the proton
%     (4-momentum $P$).
%     The gluon carries a fraction $\xi$ of the (longitudinal) 
%     proton 4-momentum. The virtual quark entering the I-subprocess
%     has 4-momentum $q'$,
%     and the outgoing quark ({\it = current quark}) from the 
%     $\gamma \rightarrow q \ol q $-process
%     has 4-momentum $q^{\prime\prime}$. $x'$ is formally defined 
%     like $x_{\rm Bj}$, but cannot be illustrated in a simple way.
%     $0 \leq x_{\rm Bj} \leq x_{\rm Bj}/\xi \leq x' \leq 1$ holds. 
%     $W_i$ is the invariant mass of the quark gluon($q'g$) system
%     and $W$ is the invariant mass of the total hadronic system 
%     (the $\gamma P$ system).
 %    \shat\ is the invariant mass squared of the $\gamma g$ system.}
   \label{kin-var}
\end{figure}

In deep-inelastic $e^\pm P$ scattering, I-induced events
are predominantly associated with a process structure as sketched in
Fig.~\ref{kin-var}: A photon splitting into a quark-antiquark pair, fuses with
a gluon from the proton in the background of an
instanton or an anti-instanton. Besides the
current-quark (jet), the partonic final state consists of 
$2n_f-1$  right (left)-handed (massless) quarks and anti-quarks and
$n_g$ gluons emitted by the instanton
(anti-instanton). Correspondingly, the total
violation of {\it chirality} $ \equiv \#q_R -\#q_L$
is $\pm 2n_f$, in agreement with the ABJ chiral anomaly
relation~\cite{adler,belljarckiw,bardeen}. 

%%%%%%%%%%%%%%%%%%%%%%%%%%%%%%%%

I-induced processes initiated by a quark from the proton are
suppressed by a factor of $\alpha_s^2$ with respect to the gluon initiated
process~\cite{RS4}. This fact together with 
the high gluon density in the relevant kinematical domain at HERA,
justifies to neglect  quark initiated processes.
\vspace{0.5cm}

QCDINS~\cite{qcdins1,qcdins3} is a Monte Carlo package for simulating 
QCD-instanton induced scattering processes in DIS. It is designed as an 
``add-on'' hard process generator interfaced by default to the Monte
Carlo generator HERWIG~\cite{herwig}. 

QCDINS incorporates the
essential characteristics, that have been derived 
theoretically for the hadronic final state of I-induced processes:
notably, the {\it isotropic} production of the partonic final
state in the I-rest system ($q^\prime g$ CMS in Fig.~\ref{kin-var}), 
flavour ``democracy'', energy
weight factors different for gluons and quarks, and  a high average
multiplicity $2n_f+{\cal O}(1/\alpha_s)$ of produced partons with a 
(approximate) Poisson distribution of the gluon multiplicity.  

Let us briefly summarize the main stages involved in QCDINS, to generate
the complete I-induced partonic final state in DIS.
 
The first stage is the generation of the various Bjorken variables   
(c.\,f. Fig.~\ref{kin-var}) of the I-induced DIS process. They are 
generated in the order $Q^{\prime 2},x^\prime,\xi,x_{\rm Bj},
y_{\rm Bj}$, with weights corresponding to the I-induced total 
cross section~\cite{RS4,moch2},
\begin{equation}
\label{evwgt}
d\sigma_{eP}^{(I)} \simeq 
\frac{2\,\pi\,\alpha^2\,e_{q^\prime}^2}{S}
\ dQ^{\prime 2}\,
\frac{dx^\prime}{x^\prime}\,
\,\frac{\sigma^{(I)}_{q^\prime g}(x^\prime ,Q^{\prime 2})}{x^\prime}
\ \frac{d\xi}{\xi}f_g (\xi )\ 
\frac{dx_{\rm Bj}}{x_{\rm Bj}}\, 
\frac{dy_{\rm Bj}}{y_{\rm Bj}}\,
\frac{1+(1-y_{\rm Bj})^2}{y_{\rm Bj}}\  
P_{q^\prime}^{{ (I)}}\ 
 ,
\end{equation}
subject to appropriate kinematical restrictions  
(e.\,g. $0 \leq x_{\rm Bj} \leq x_{\rm Bj}/\xi \leq x' \leq 1$)
and fiducial cuts. Here $e_{q^\prime}^2$ denotes the electric charge 
squared of the virtual quark $q^\prime$ in units of the electric charge
squared, $e^2=4\pi\alpha$, and $f_g$ is the gluon density in the proton.  
The weight factor $P_{q^\prime}^{{ (I)}}$ accounts for the flux of 
virtual quarks $q^\prime$ in the I-background~\cite{RS4,moch2},
\begin{eqnarray}
P_{q^\prime}^{{ (I)}} = \frac{3}{16\,\pi^3}\,\frac{x_{\rm Bj}}{\xi\, x^\prime}
\left(1+\frac{\xi}{x_{\rm Bj}}
-\frac{1}{x^{\prime}}-\frac{Q^{\prime 2}}{S x_{\rm Bj}y_{\rm Bj}}\right)
\, .
\end{eqnarray}

\begin{figure} [ht]
\centering
\mbox{
\epsfig{file=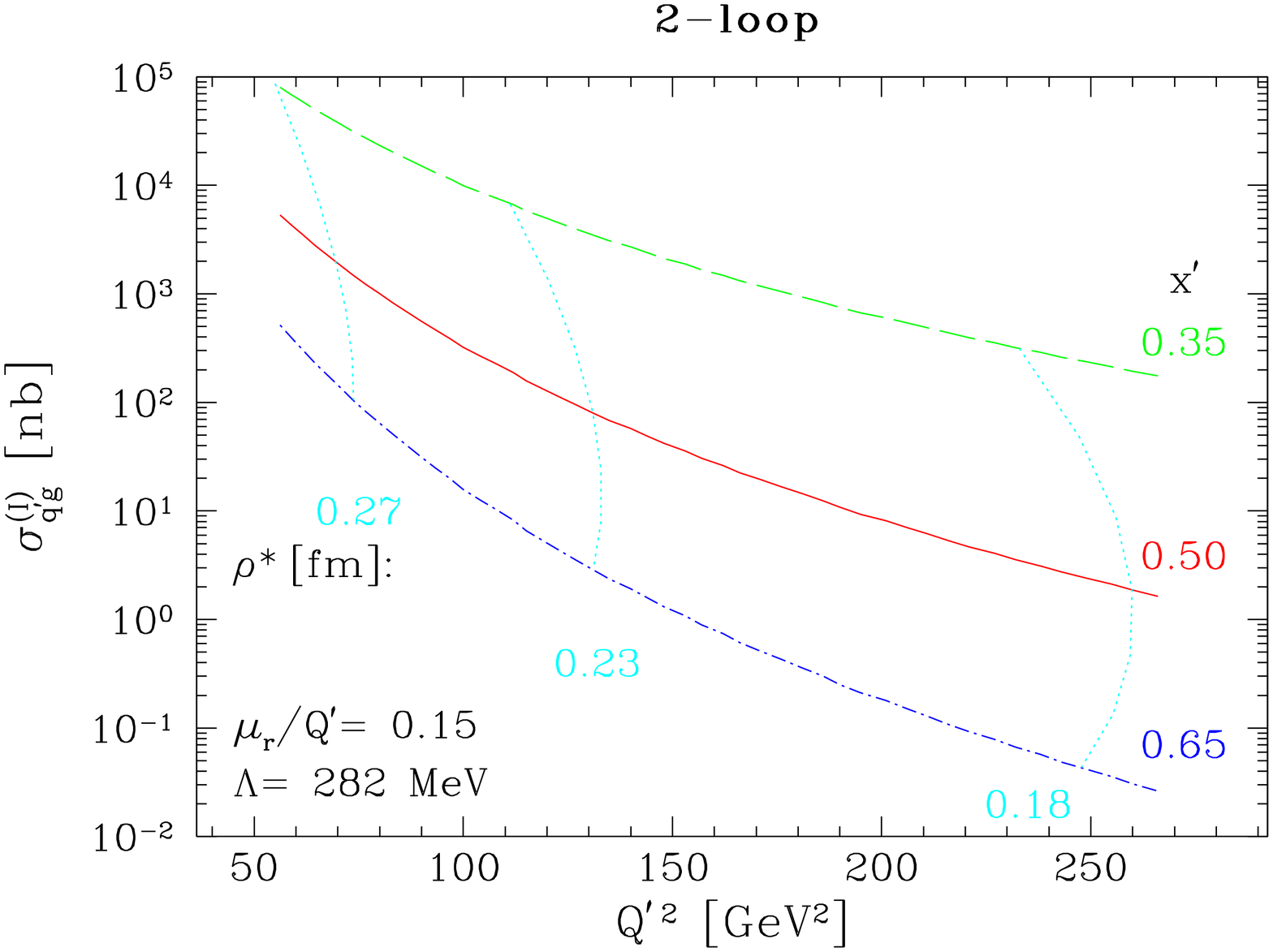,angle=-90,width=8.3cm}
\epsfig{file=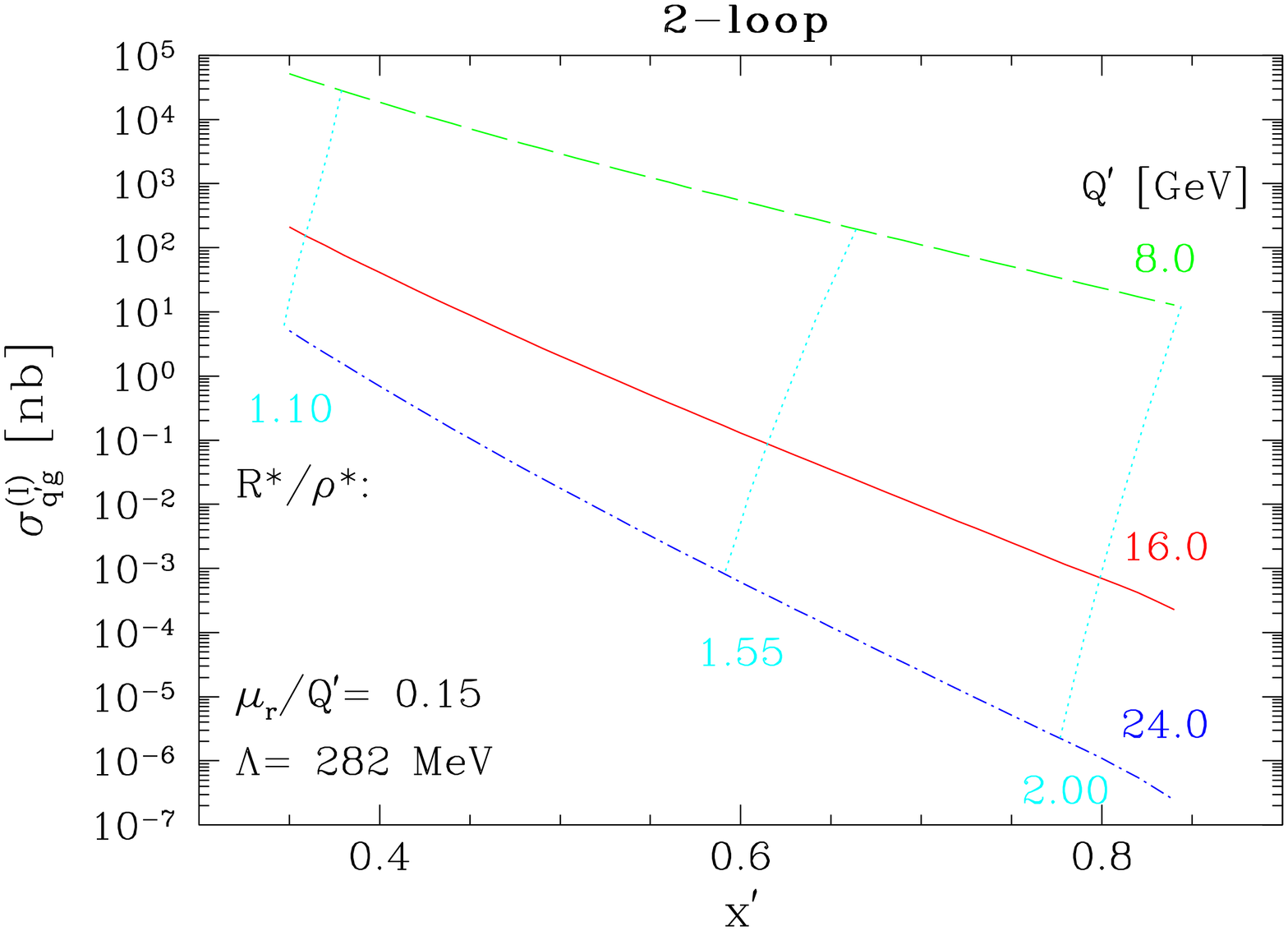,angle=-90,width=8.3cm}
     }
  \caption[]
   {I-subprocess cross section~\cite{RS4} displayed versus 
   the Bjorken variable $Q^{\prime\,2}$ with
   $x^\prime$ fixed
   (left) and versus $x^\prime$ with $Q^{\prime\,2}$ fixed (right) for 
    $n_f=3$. The dotted lines indicate the corresponding effective
    $I$-sizes $\rho^\ast$ [fm] (left) and $I\overline{I}$-distances
    $R^\ast$ in units of $\rho^\ast$ (right), respectively.}
  \label{isorho}
\end{figure}

In Eq.~(\ref{evwgt}), the I-induced $q^\prime g$-subprocess total 
cross section $\sigma_{q^\prime g}^{(I)}$ contains the essential 
instanton dynamics~\cite{RS4}. As illustrated in Fig.~\ref{isorho}, 
it is very steeply growing 
%%%%%%%%%%%%%%%%%%%%%%%%%%%%%FIGURE  %%%%%%%%%%%%%%%%%%%%%%%%%%%%%%%%%%%5
%%%%%%%%%%%%%%%%%%%%%%%%%%%%%%%%%%%%%%%%%%%%%%%%%%%%%%%%%%%%%%%%%%%%%%
for decreasing values of $Q^{\prime 2}$ and $x^\prime$, respectively. 
In order to remain within the region of validity of I-perturbation 
theory~\cite{RS4,rs-lat}, the following cuts are implemented by default
in QCDINS,
\begin{equation} 
\label{fiducial}
Q^{\prime 2}\geq Q^{\prime 2}_{min} = 64\ {\rm GeV}^2 \, ; \hspace{6ex}
x^\prime \geq x^\prime_{min} = 0.35 .
\end{equation}
An additional cut on the photon virtuality,
\begin{equation}
\label{fiducialQ}
S x_{\rm Bj} y_{\rm Bj} \equiv Q^2 \geq Q^2_{min} = 64\ {\rm GeV}^2 ,
\end{equation}
should in principle be applied in order to warrant sufficient suppression of 
non-planar contributions~\cite{moch}, which may spoil the  
validity of Eq.~(\ref{evwgt}).
The cross section $\sigma_{q^\prime g}^{(I)}$ depends strongly on
the QCD scale  $\Lambda^{(n_f)}_{\overline{\rm MS}}$ and weakly on
the renormalization scale $\mu_r$. By default, these scales are taken as
\begin{equation}
\Lambda^{(3)}_{\overline{\rm MS}}= 0.282\ {\rm GeV} \, ;\hspace{6ex}
\mu_r = 0.15\,Q^\prime .
\label{lambda}
\end{equation}
Note that strictly speaking, the underlying theoretical framework
refers to massless quarks. Therefore, the (default) number of flavours
was set to $n_f=3$ and $\Lambda^{(3)}_{\overline{\rm MS}}$ in
Eq.~(\ref{lambda}) was obtained by a standard flavour reduction from the
DIS average, $\Lambda^{(4)}_{\overline{\rm MS}}=234$ MeV in
Ref.~\cite{PDG}. The 
default value for $\mu_r$ above corresponds to the minimum~\cite{RS4},
$\partial \sigma_{q^\prime\,g}/\partial\mu_r = 0$.
 
In the second stage of the event generation, the 4-momenta $g,q,q^\prime,
q^{\prime\prime}$ of the incoming gluon $g$, the virtual photon $q$, 
the virtual quark $q^\prime$ and the current quark 
$q^{\prime\prime}$, respectively, are filled. Sudakov
decompositions of these momenta are used to incorporate various
constraints, e.g. on the momenta squared, 
$g^2=m_g^2(=(0.75\  {\rm GeV})^2$ by default), $q^2=-Q^2$, $q^{\prime 
2}=-Q^{\prime2}$, $q^{\prime\prime 2}=m_{q^{\prime\prime}}^2$. The 4-momentum
of the outgoing lepton, $e^\prime$, is calculated subsequently. 

In the third stage, the partonic final state of the I-induced 
$q^\prime g$-subprocess is generated in the $q^\prime g$ CMS as follows.
The number $n_g$ of produced gluons is generated according to 
a Poisson distribution with mean $\langle n_g\rangle^{(I)} \sim
1/\alpha_s\sim 3$ as calculated theoretically in I-perturbation
theory. $n_f (=3)$  $[\,q \ldots \overline{q}\,]$ - ``strings'' of
partons are set up, each beginning  
with a quark, 
followed by a random number of gluons and ending with an anti-quark of
randomly chosen flavour. There are $n_g+1$ gluons in total and all
$n_f$ flavours occur precisely once.   
A simple example for $n_g=3$ outgoing gluons may provide some illustration:
$[\,u\; g\; g\;\overline{s}\,]$ $[\,d\;g\;\overline{u}\,]$
$[\,s\;g\;g\;\overline{d}\,]$. 
A quark and a gluon among these $2n_f+n_g+1$ partons are (randomly) marked
as incoming.

The momenta $p_i$ of the $n=2n_f-1+n_g$ outgoing partons in the instanton 
subprocess CMS are then generated according to the 
energy-weighted phase-space distribution
\begin{eqnarray}
\label{leadingorder}
\lefteqn{\int 
\prod_{i=1}^{2n_f-1}
\left\{ d^4p_i\,\delta \left( p_i^2-m_i^2\right) p^0_i\right\}
\prod_{k=1}^{n_g}
\left\{ d^4p_k\,\delta \left( p_k^2-m_g^2\right) p^{0\,2}_k\right\}}
\\ \nonumber &&\times
\, \delta^{(3)}\left( \sum_{i=1}^{2n_f-1} \vec{p}_i+
\sum_{k=1}^{n_g} \vec{p}_k \right)\,
\delta \left( W_i - \sum_{i=1}^{2n_f-1} p_i^0 - 
\sum_{k=1}^{n_g} p_k^0\right) , 
\end{eqnarray} 
corresponding to the leading-order matrix element with different
energy weights for gluons and quarks.

Next, the colour and flavour connections of the partons are set up. The
colour flow is obtained  simply by 
connecting the colour lines of neighbouring partons within each of the
above-mentioned $n_f$ $[\,q \ldots \overline{q}\,]$ - ``strings'' in  a
planar manner (consistent with the leading order $1/N_c$ expectation).  

The partonic stage of the event generation ends by boosting the momenta of
the I-subprocess final state partons to the laboratory frame. 

After the perturbative evolution of the generated partons, one may use 
the hadronization mechanisms inherent in various Monte Carlo models (e.\,g. 
HERWIG~\cite{herwig} or JETSET~\cite{string3}) to arrive at the
complete hadronic final state.

\section{Cross section and Signature \label{signature}}

The total cross section of I-induced events in DIS, calculated within
I-perturbation theory, is
surprisingly large. For $x_{\rm Bj} \geq 10^{-3}$ and $0.1 < y_{\rm Bj} < 0.9$,
the result of Ref.~\cite{RS4} is
\begin{equation}
\sigma_{HERA}^{(I)}(x' \geq 0.35; Q' \geq 8 ~\mbox{GeV})~\simeq~ 
                     130 \,\mbox{pb}.
\label{WQhera}
\end{equation}
Hence, given the total integrated luminosity $\mathcal{L} \simeq
30\,\mbox{pb}^{-1}$ accumulated by each of the 
HERA experiments in the years 1996/1997, 
in this kinematical region a large number  
$N= \sigma_{HERA}^{(I)} \cdot \mathcal{L}=\order{10^4}$ of
I-induced events should already have  been taken on tape.

The cross section (\ref{WQhera}) corresponds to a fraction of I-induced
to normal DIS (nDIS) events of~\cite{RS4} 
\begin{equation}
f^{(I)} ~=~ \frac{\sigma_{HERA}^{(I)}}{\sigma_{HERA}^{(nDIS)}}~=~\order{1} \%.
\label{fraction}
\end{equation}
However, this prediction for the cross section 
still contains considerable uncertainties~\cite{RS4}. One of the
dominant ones arises from the experimental uncertainty of $\pm 65$ MeV 
associated with the QCD scale $\Lambda^{(3)}_{\overline{\rm MS}}$ in 
Eq.~(\ref{lambda}). If $\Lambda_{\overline{\rm MS}}^{(3)}$ is varied 
within the allowed range of $\pm 65$ MeV,
the cross section varies by $\order{+300}$\,pb and
$\order{-100}$\,pb, respectively.

By far the dominant uncertainty, however, comes in principle from 
the choice of the lower bounds on $x'$ and $Q'$, due to the strong  
increase of the I-subprocess cross section with
decreasing values of the cuts $x'_{min}$ and $Q'_{min}$, respectively
(cf. Fig.~\ref{isorho}). 
However, recent high-quality ``data'' from a (non-perturbative) lattice
simulation~\cite{teper} on the topological structure of the QCD vacuum  
have strongly reduced this uncertainty.
The lattice results could be directly converted into
a ``fiducial'' $(x^\prime, Q^\prime)$ region, where the predictions from 
I-perturbation theory should (approximately) hold~\cite{rs-lat}. 
The lower bounds on  $x'$ and $Q'$ given 
in Eq.~(\ref{WQhera}) correspond to this fiducial region and are thus
expected to be quite reliable . 
%
%%%%%%%%%%%%%%%%%%%%%%%%%%%%%%%%%%%%%%%%%%%%%%%%%%%%%%%%%%%%%%%%%%%%
\begin{figure} [ht!]
\centering
\mbox{
   \epsfig{file=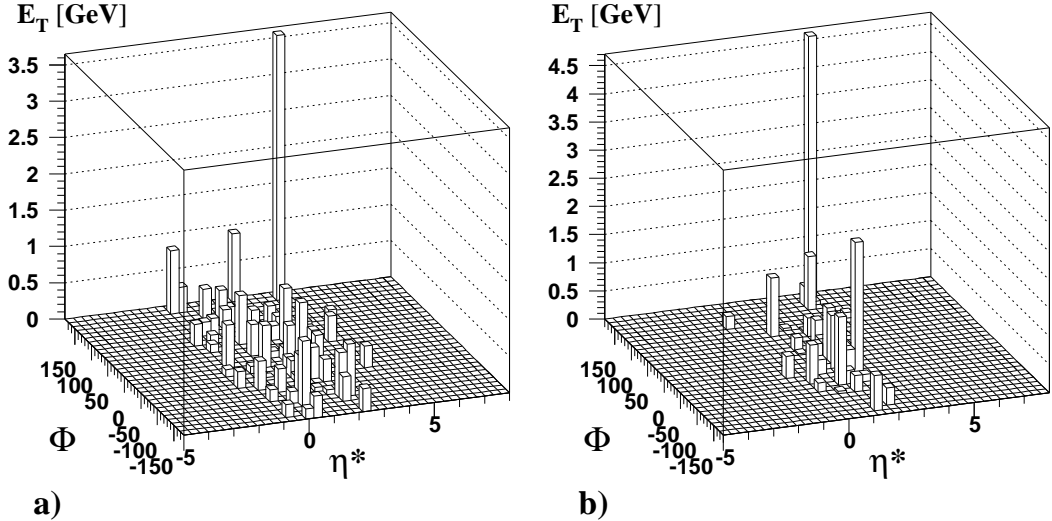,
      width=14cm,bbllx=20pt,bblly=236pt,bburx=575pt,bbury=610,clip=}
    }
   \caption[]
     {Transverse energy distributions in the $\eta^*$-$\Phi$-plane
      (hCMS) of the hadronic final state particles of an 
     I-induced process (%the regions $\eta^* \rightarrow \pm \infty$ 
     shown for typical detector acceptance cuts).
     a) ``Ideal'' event signature. Besides the homogeneously populated 
     ``instanton-band'' (at $\eta^*$ between 0 and 2), the current jet at 
     $\eta^* \approx 3$, $\Phi \approx 170^\circ$ is clearly visible.
     b) The homogeneous distribution of the hadrons in the instanton band
     is destroyed by a current jet with high $p_T$.}
   \label{legos}
\end{figure}
%%%%%%%%%%%%%%%%%%%%%%%%%%%%%%%%%%%%%%%%%%%%%%%%%%%%%%%%%%%%%%%%%%%%%%%

Let us next turn to the expected signature of the I-induced final
state.

Fig.~\ref{legos}a shows the transverse energy distribution in the
$\eta^*$-$\Phi$ plane of a ``typical'' I-induced event
in DIS, produced by the instanton Monte Carlo generator 
QCDINS~\cite{qcdins1,qcdins3}.
There are the following characteristics:
%%%%%%%%%%%%%%%
\begin{itemize}
\item {\bf Topology: ``Instanton band'' and current jet}\\
In the instanton CMS (the $q'g$ CMS, cf. 
Fig.~\ref{kin-var}) the production
of partons coming from the I-subprocess
is isotropic.
This leads to an energy distribution of
hadrons restricted to a certain range in the pseudo-rapidity 
$\eta$ (the ``instanton-band'') with a bandwidth of
$\simeq \pm 1.1$ units in $\eta$. 
In the instanton CMS this band is localized around 
$\eta = 0$, while in the hCMS, the center of the band is
shifted to higher values (depending on the kinematic variables
$x_{\rm Bj}$ and $Q^2$).
Besides the band, the hadronic final state of I-induced events 
exhibits a current jet coming from
the outgoing current quark (denoted by $q''$ in Fig.~\ref{kin-var}).
%%%%%%%%%%%%%%%%%%%%%%%%%%%%%%%%%
\item {\bf High multiplicity and ``Flavour-democracy''}\\
In every I-induced event, one pair of quark and anti-quark 
of all $n_f(=3)$ flavours is simultaneously produced.
In addition, a mean of $\langle n_g \rangle ^{(I)} \sim  1/\alpha_s  \sim  3$
gluons is expected.
Hence, for the phase space studied here, we find a mean number of 
partons of $\order{10}$, leading to a mean multiplicity of charged
particles of $\order{20}$ in every event.
The actual number of hadrons produced mainly depends on the center 
of mass energy $W_i$ accessible.
Moreover, due to the democratic production of all (effectively massless)
flavours, more mesons containing heavy quarks should be found in 
I-induced events than in normal DIS events. The detection of
$K^0$ mesons is experimentally most promising to exploit this feature.
%%%%%%%%%%%%%%%%%%%%%%%%%%%%%%%%%%%%
\item {\bf High transverse energy}\\
While normal DIS events exhibit a mean value of 2 GeV of transverse
energy per unit of $\eta^*$ \cite{daten1},
for I-induced events a value of the order of 5 GeV % (for small values of $x'$) 
in an $\eta^*$ range of $0 \lsim \eta^* \lsim 4$ is expected.
\end{itemize}

\begin{figure} [htb]
\centering
\mbox{
   \epsfig{file=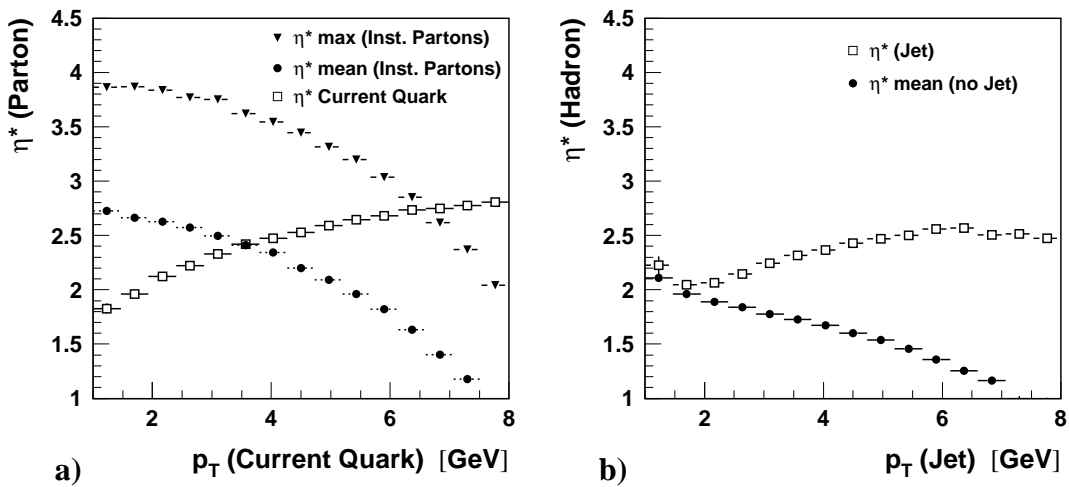,
      width=14cm,bbllx=25pt,bblly=260pt,bburx=565pt,bbury=510,clip=}
    }
   \caption[ ]
     {Evolution of the relative $\eta^*$ position of the partons belonging
      to the $I$-subprocess and the current jet 
     % the topology of I-induced events in terms of pseudorapidity in 
     as a function of the $p_T$ of the current quark
      and the current jet, respectively.
%   mean $\eta^*$-position of the ``instanton-band'' and
%      the current quark (a) and the current jet (b) in function of
%      the $p_T$ of the current quark and the current jet, respectively.
      In a) the instanton-band is outlined by the maximal and
      the $E_T$ weighted mean $\eta^*$ value of the partons coming from the 
      I-subprocess.
      In b) the position of the band is characterized by
      the $E_T$ weighted mean $\eta^*$ value (calculated without the hadrons
      belonging to the jet).}
   \label{eta-ptjet}
\end{figure}

\noindent
The ideal event topology described above (see also Fig.~\ref{legos}a)
is quite often modified by simple kinematic 
effects. In the hCMS the sum of the transverse energy of the particles
emerging from the I-subprocess and the $p_T$ of the current
jet has to be approximately balanced. In many cases this leads to
a destruction of the homogeneous $\Phi$ distribution of the particles 
in the band.
Fig.~\ref{legos}b gives an example for such an event topology.
In general, with rising $p_T$ of the current jet and 
lower particle multiplicities emerging from the
I-subprocess, the band structure becomes less clearly visible.\\
In addition, the $p_T$ of the jet strongly influences the
relative $\eta^*$ position of the jet and the instanton band. 
Fig.~\ref{eta-ptjet}a demonstrates this effect for the parton final state. 
The mean $\eta^*$ value of the current quark increases
with rising $p_T$ of the current quark, while the mean $\eta^*$ values of the 
instanton band,
outlined by the maximal and the mean ($E_T$ weighted) $\eta^*$ value
of the partons coming from the I-subprocess, are decreasing.
For hadrons (Fig.~\ref{eta-ptjet}b) the $\eta^*$ position of 
the current jet (which is reconstructed by a jet-finding algorithm, see
section \ref{recon}) and the $E_T$ weighted mean $\eta^*$ value of 
all particles not belonging to the jet 
(which we use as an estimator for the central
position of the instanton band) show qualitatively the same 
behaviour\footnote{The mean $\eta^*$ values of the centre of the 
instanton band are lower than those for the parton final state, 
because for hadrons,
particles of the proton remnant are also taken into account.
Additionally, the mean $\eta^*$ values for 
$p_T(current\ quark)\lsim 3~\mbox{GeV}$ are
not adequately reproduced on hadron level. In this range, the jet-finding
algorithm mostly finds the ``jet'' within the hadrons belonging to
the instanton band.}.
%%%%%%%%%%%%%%%%%%%
%%%%%%%%%%%%%%%%%%%%%%%%%%%%%%%%%%%%%%%%%%%%%%%%%%%%%%%%%%%%%%%%%%%%%%%

In Fig.~\ref{eta-ptjet} and throughout this report, the
kinematical range given
in the context of Eq.~(\ref{WQhera}) has actually been extended 
down to $x_{\rm Bj}=10^{-4}$ with $Q^2=
Sx_{\rm Bj}y_{\rm Bj}\geq 5 ~\mbox{GeV}^2$. On the one hand, within this 
enlarged kinematical region, there is the possible influence of non-planar 
graphs, which are not implemented in QCDINS. 
On the other hand, the hadronic final state topology of I-induced events
will presumably not change dramatically, 
being mainly determined by the available phase space (see section 
\ref{depend}).
Since the cross section of I-induced events increases
with decreasing values of $x_{\rm Bj}$ and $Q^2$ (as for normal DIS events), 
it is an experimental challenge to investigate the signal to background
ratio also in this kinematical domain. In our  extended study in 
Ref.~\cite{qcdins2} we shall restrict ourselves to the ``fiducial'' 
high-$Q^2$ regime~(Eq. \ref{fiducialQ}). 

%%% Local Variables: 
%%% mode: latex
%%% TeX-master: "kinem"
%%% End: 
%%%%%%%%%%%%%%%%%%%%%%%%%%%%%%%%%%%%%%%%%%%%%%%%%%%%%%%%%%%%%%%%%%%%%%%%%
\section{Reconstructing the kinematics of the I-subprocess\label{recon}}
In this section we give a short summary of the possibility to reconstruct
the kinematic variables of the I-subprocess which have been introduced
in Fig.~\ref{kin-var}. A more detailed
investigation can be found in \cite{jgerigk}.
The cross section of the I-subprocess is most sensitive to the values
of ${Q'}^2$ and $x'$. Therefore the possibility to reconstruct 
theses variables would
allow crucial tests of the predictions for the total
cross section to be made, once I-induced topologies have been 
identified.
Moreover, in order to reconstruct ${Q'}^2$ and $W_i$, the separation 
of the current jet from the instanton band has to be as good as possible.
We found the output of this procedure to be most useful in designing
optimized observables in view
of enhancing the separation power to normal DIS events.
And finally, the reconstruction of ${Q'}^2$ and $\xi$ would allow
a boost to the instanton center of mass frame to be performed.

%%%%%%%%%%%%%%%%%%%%%%%%%%%%%%%%%%%%%%%%%%%%%%%%%%%%%%%%%%%%%%%%%%%%%%%
\begin{figure} [ht!]
\centering
\mbox{
   \epsfig{file=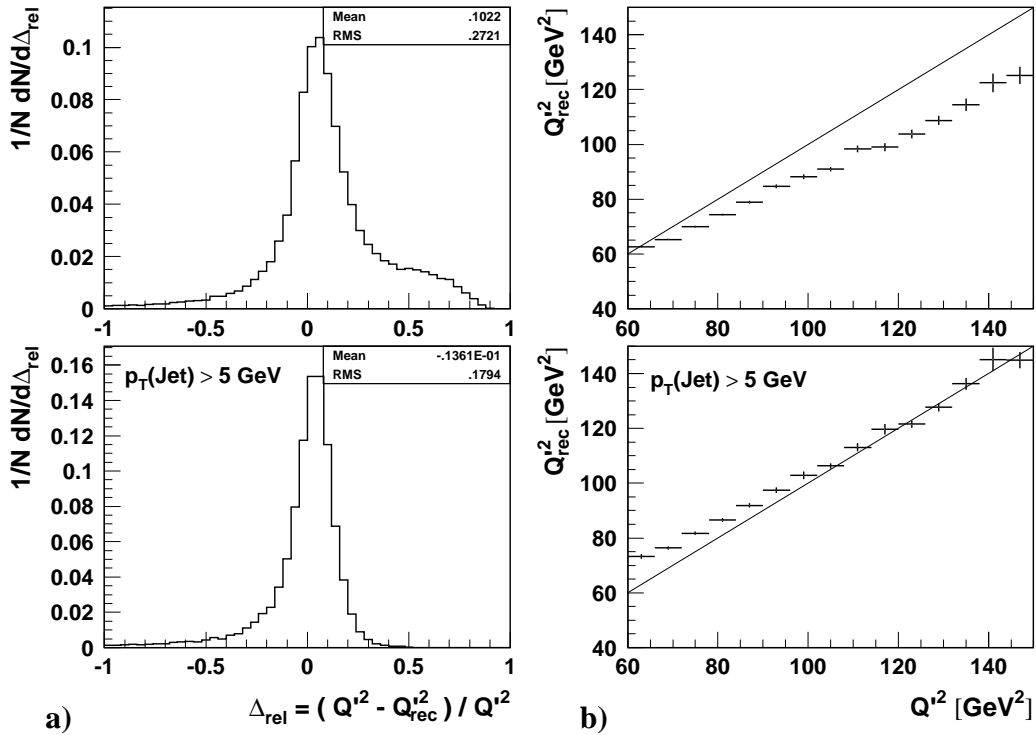,
      width=14cm,bbllx=20pt,bblly=210pt,bburx=570pt,bbury=600,clip=}
    }
   \caption[]
     {Reconstruction of ${Q'}^2$ from the hadronic final state of I-induced
      events. a) Relative deviation of the true and the reconstructed value.
      b) Correlation between the mean value of the reconstructed ${Q'}^2$
      and the true value. In both cases the upper figure is without
      and the lower figure is with a $p_T$ cut on the current jet.}
%, without (upper Fig.) and with (lower Fig.) a cut on the $p_T$ of
% the current jet.     
%true value, again without and with $p_T$ cut on the current jet.}
   \label{qprime-rec}
\end{figure}
%%%%%%%%%%%%%%%%%%%%%%%%%%%%%%%%%%%%%%%%%%%%%%%%%%%%%%%%%%%%%%%%%%%%%%%

The reconstruction of ${Q'}^2$ is based on the assumption 
%at the $\gamma \rightarrow q \ol{q}$ vertex four-momentum conservation
%holds, i.e. 
that the colour forces between the current jet and the rest
of the partons,
%of the particles 
as modeled in the hadronisation phase,
still allow the reconstruction of the current jet 4-vector.
% can be neglected.
So $q'=q-q''$ (cf. Fig.~\ref{kin-var}) and
the reconstruction of ${Q'}^2$ is performed by reconstructing the four-vector
$q''$ of the particles belonging to the current jet. To identify 
the current jet the cone algorithm \cite{pozo,cdfcone} is applied in
the hCMS. A cone of radius $R=0.5$ (in the $\eta^*$-$\Phi$-plane)
turns out to perform best \cite{jgerigk}. 
Motivated by the relative $\eta^*$ position of the current quark and
the instanton band (cf. Fig.~\ref{eta-ptjet}),
we additionally require the jet to fulfill
$\eta^* \geq \ol{\eta^*}$,
with $\ol{\eta^*}$ being the $E_T$-weighted mean $\eta^*$ value of all hadrons.
The jet with the highest $p_T$ is then assumed to be the most likely
candidate for the current jet. Fig.~\ref{qprime-rec} shows that this
procedure gives a rather good reconstruction of ${Q'}^2$, especially, if
the jet is required to have a $p_T \geq 5 ~\mbox{GeV}$.
This is also a reasonable choice in order to have a clear signal 
in the detector.\\
Once the current jet has been found, the hadrons belonging to the
jet are removed from the final state
and $\ol{\eta^*}$ is recalculated with the remaining hadrons
in order to get a good estimator for the center of the instanton band.
For the calculation of $W_i$ the four-vectors of all hadrons in a region
left and right of $\ol{\eta^*}$ are considered. A band of
~$\ol{\eta^*} \pm 1.1$ gives the best possible results 
for the majority of events.
This result is in agreement with the mean bandwidth
visible in Fig.~\ref{eta-ptjet} as well as with the
expectation for an isotropic event. 
The reconstruction of $W_i$ is not perfect, though. If the bandwidth
is determined for every single
event (based on the known value of $W_i$), a rather broad distribution is
obtained, which peaks only weakly at $\pm 1.1$. In many events, slightly
different bandwidths would be required for an optimal reconstruction
of $W_i$. 
%We did not find any obvious event property to get control on
%the actual bandwidth required. 
%This results 
%
The procedure using the averaged bandwidth results 
in a quite large error of the reconstructed
quantity of $W_i$ of approx. 25 \% (for $p_T(jet) \geq 5 
~\mbox{GeV}$).\\

%Since the bandwidth required seems mainly to be determined by
%fluctuations in the hadronisation phase, on which you have no control, this
%results in a quite large error of the reconstructed
%quantity of $W_i$ of approx. 25 \% (for $p_T(jet) \geq 5
%~\mbox{GeV}$).\\ %
%Since the I-subprocess is fully determined by two kinematic variables, 
The variable $x'$ can be calculated once ${Q'}^2$ and $W_i$
have been reconstructed. However, the errors 
of the ${Q'}^2$ and the $W_i$ reconstruction interact in such an unfavourable
manner that a reconstruction of $x'$ seems not very meaningful.
Finally,
$\xi$ can be reconstructed via \shat, which comprises the four-vectors of
all the hadrons used for the reconstruction of ${Q'}^2$ and $W_i$. 
The reconstruction of $\xi$ works relatively well in a broad range of
the variable. However, a reconstructed pair of $q'$ and $\xi$ leads to
inconsistencies when trying to perform a boost to the instanton center of
mass system in $50\%$ of the cases, so that either variable would have to
be corrected.
%
%

%%% Local Variables: 
%%% mode: latex
%%% TeX-master: t
%%% End: 
\section{Search Strategies \label{search}}
\begin{figure} [ht!]
\centering
\mbox{
   \epsfig{file=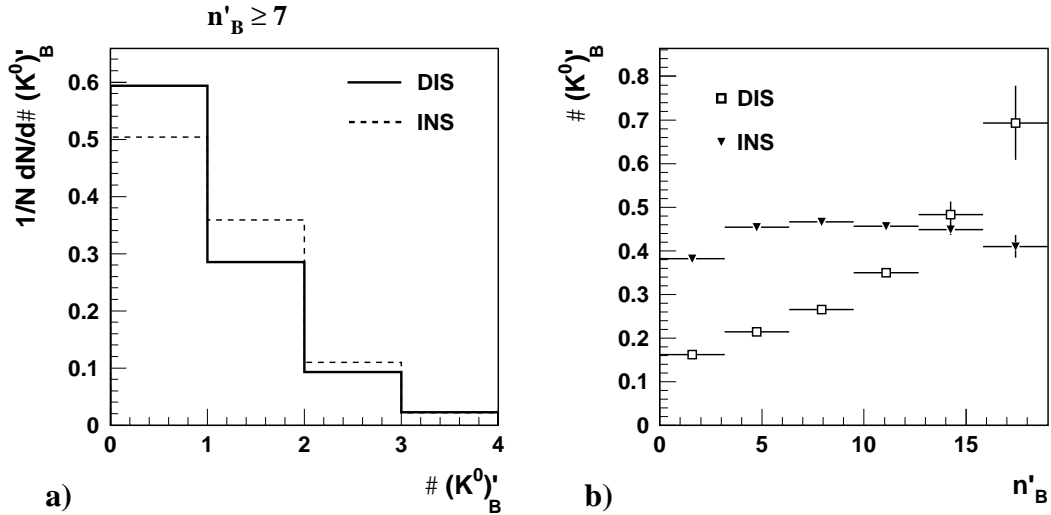,
      width=14cm,bbllx=25pt,bblly=270pt,bburx=570pt,bbury=535,clip=}
    }
   \caption[]
    {a) Shape normalized distributions of the number of neutral
    kaons in the instanton band for normal DIS and I-induced events, 
    while applying the additional cut $n'_B \ge 7$.
    b) Correlation between the mean number of neutral kaons and the
    number of charged particles.}
   \label{kaon-cor}
\end{figure}

A search strategy for I-induced processes is 
naturally based on the characteristic properties of the 
hadronic final state.
The goal is to isolate - based
on Monte Carlo predictions - an instanton enriched data sample by applying
cuts on a set of observables.
More than 30 observables have been investigated
%in this respect 
in \cite{jgerigk}\footnote{%In an extension of the cited paper,
%the results presented in this report are based on a study
%which uses additional cuts 
For the results shown here additional cuts have been introduced
to get a more realistic 
representation of the typical acceptance of a HERA detector.
For this reason, some of the results presented here slightly deviate from those
given in \cite{jgerigk}.}.
%%%%%%%%%%%%%%%%%%%%%%%%%%%%%%%%%%%%%%%%%%%%%%%%%%%%%%%%%%%%%%%%%%%%%%%%%
\begin{figure} [ht!]
\centering
\mbox{
   \epsfig{file=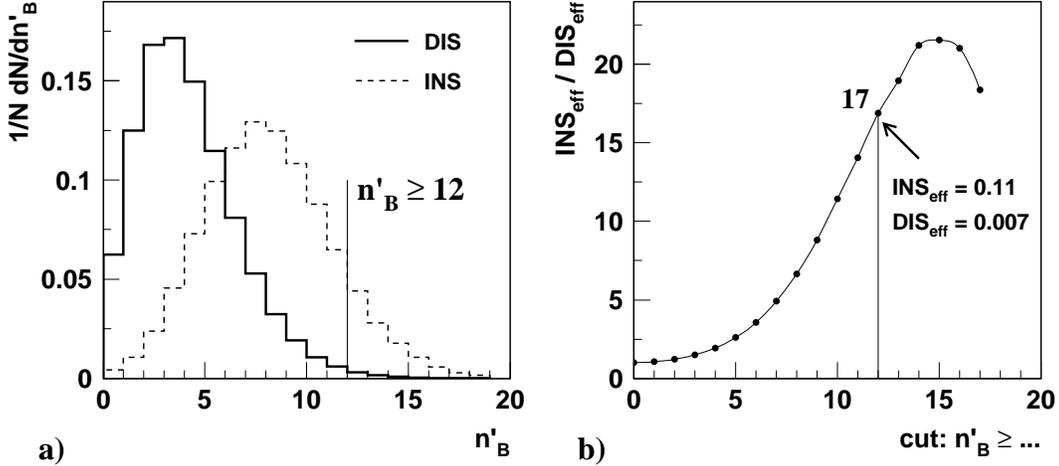,
      width=14cm,bbllx=25pt,bblly=270pt,bburx=570pt,bbury=515,clip=}
    }
   \caption[]
           {a) Shape normalized distributions of the number of charged
    particles in the instanton band $n'_B$
   (without the hadrons belonging
    to the current jet) for normal DIS and I-induced events.
    b) Separation power (:= $INS_{eff}/DIS_{eff}$) after applying cuts
    in $n'_B$ for every possible cut. Explicitly
    shown are the values for the cut with the highest separation power, 
    which fulfills the requirement of a remaining instanton efficiency
    of 10 \%.}
   \label{nb-onedim}
\end{figure}
%%%%%%%%%%%%%%%%%%%%%%%%%%%%%%%%%%%%%%%%%%%%%%%%%%%%%%%%%%%%%%%%%%%%%%%
Fig.~\ref{nb-onedim} demonstrates the procedure by which we
measure the ``separation power'' of each observable. As an example,
we investigate the distribution of the number of charged particles $n'_B$,
where the index ``$B$'' indicates that we consider only particles of the
instanton band ($\ol{\eta^*}\pm 1.1$), and the prime indicates that
the hadrons belonging to the current jet are not considered.
Starting from shape normalized distributions for normal DIS and
I-induced events (Fig.~\ref{nb-onedim}a), we calculate the
ratio of efficiencies 
for I-induced over normal DIS events for each possible cut value 
(Fig.~\ref{nb-onedim}b). We require a minimum
instanton efficiency, which we (arbitrarily) set to 10 \%.
The cut $n'_B \ge 12$ leads to a ratio of the efficiencies, or - in
our terminology - to a separation power of $INS_{eff}/DIS_{eff}\simeq 17$.
In general, applying cuts in only {\it one} observable 
typically leads to a separation power not higher than \order{20}
(c.\,f. \cite{jgerigk}).\\
An obvious improvement in separating I-induced from 
normal DIS events is obtained by the combination of cuts in several 
observables.
However, since most of the observables are highly correlated, the
naive combination of the best one-dimensional cuts quite often fails
in enhancing the separation power. On the contrary, in some cases
it would even lead to a reduction of the separation power, an effect for
which we give an example now.
Since in any I-induced event at least one $s\ol{s}$ pair is produced,
an excess in the number of neutral kaons in comparison to normal DIS events
is expected.
Fig.~\ref{kaon-cor}a shows, however, that this excess is 
almost not visible for events with higher particle multiplicities. 
Investigating the correlation between the mean number of kaons and the number
of charged particles (Fig.~\ref{kaon-cor}b), we find 
the expected excess for relatively low particle multiplicities, only.
For some value of $n'_B$, the number of neutral kaons produced in normal
DIS events actually {\it exceeds} the number produced in I-induced 
events.
Applying cuts requiring a high value of $n'_B$
together with a large number of neutral kaons would therefore
reduce the separation power. 
Since in addition the number of kaons produced
rather strongly depends on variations of the underlying MC model,
kaons play no role in our multi-dimensional cut-scenario.
The study of strange mesons or baryons (notably $\Lambda$'s) could,
however, play a major 
role once an instanton enriched sample is experimentally isolated
from the data.\\
%%%%%%%%%%%%%%%%%%%%%%%%%%%%%%%%%%%%%%%%%%%%%%%%%%%%%%%%%%%%%%%%%%%%%%%%%
%%%%%%%%%%%%%%%%%%%%%%%%%%%%%%%%%%%%%%%%%%%%%%%%%%%%%%%%%%%%%%%%%%%%%%%

\begin{figure} [ht!]
\centering
\mbox{
   \epsfig{file=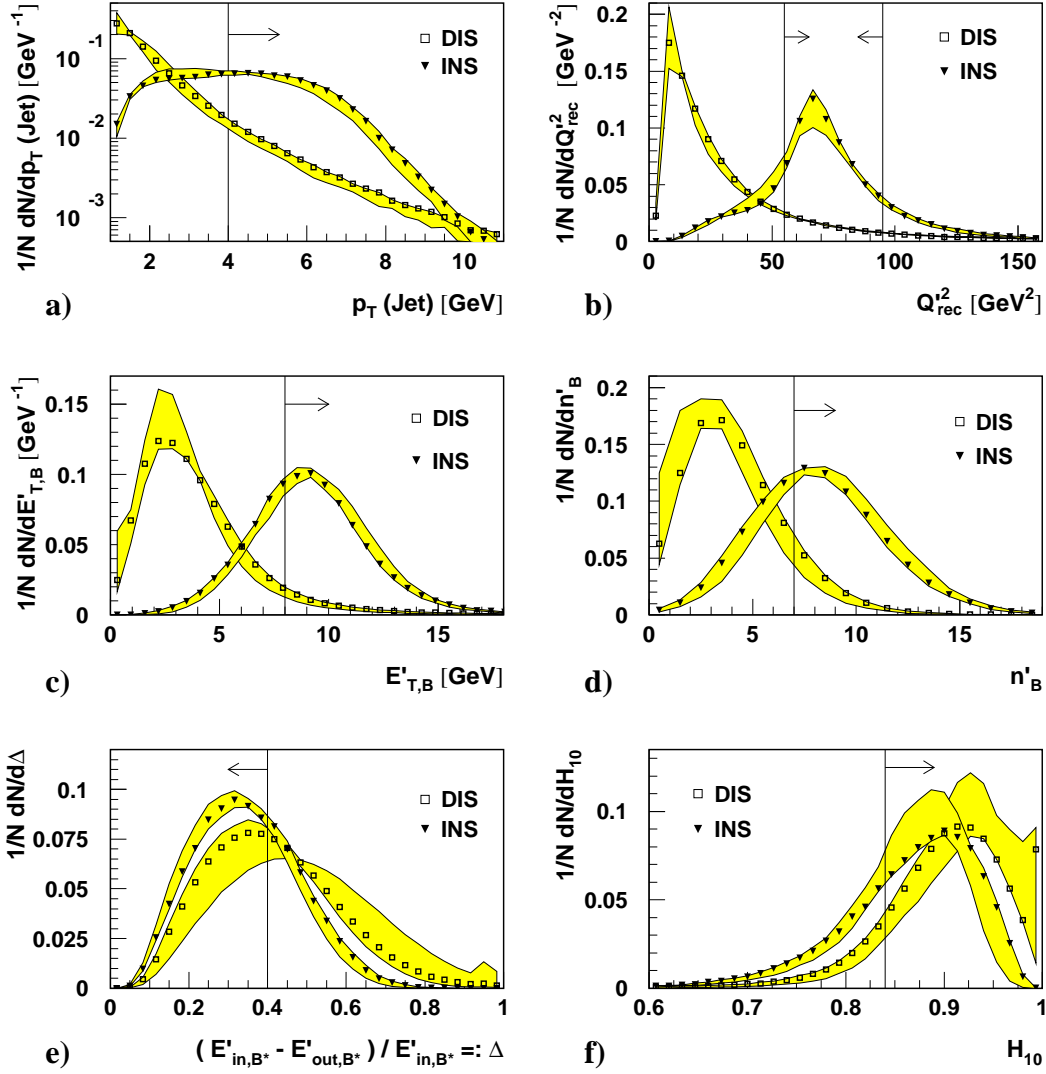,
      width=14cm,bbllx=25pt,bblly=135pt,bburx=570pt,bbury=700,clip=}
    }
   \caption[]
    {Distributions of various observables for normal DIS and 
    I-induced processes. Shown are the distributions
    for the ``reference Monte Carlos'' (INS markers = QCDINS + HERWIG,
    DIS markers = ARIADNE, including Pomeron exchange) and their variations
    (shaded band) resulting from the choice of different models or
    the variation of parameters of a model
    (c.\,f. Fig.~\ref{var-signal} and
    Fig.~\ref{var-background}). The lines and the corresponding arrows show 
    the cut applied in each of the observables, with the arrows pointing
    in the direction of the allowed region.}
   \label{6var}
\end{figure}

Based on a study of the correlations and mutual influences of
different cuts, we now introduce a set of six observables
which are able to enhance the separation power from \order{20} to 
\order{130}. Fig.~\ref{6var} shows these observables together with
the cuts applied in each observable,
indicated by the lines and the corresponding arrows\footnote{
An explicit listing of these cuts is given in Figs.~\ref{var-signal} and
\ref{var-background}.}. 
%%%%%%%%%%%%%%%%%%%%%%%%%%%%%%%%%%%%%%%%%%%%%%%%%%%%%%%%%%%%%%%%%%%%%%%
%%%%%%%%%%%%%%%%%%%%%%%%%%%%%%%%%%%%%%%%%%%%%%%%%%%%%%%%%%%%%%%%%%%%%%%

\begin{figure} [ht!]
\centering
\mbox{
   \epsfig{file=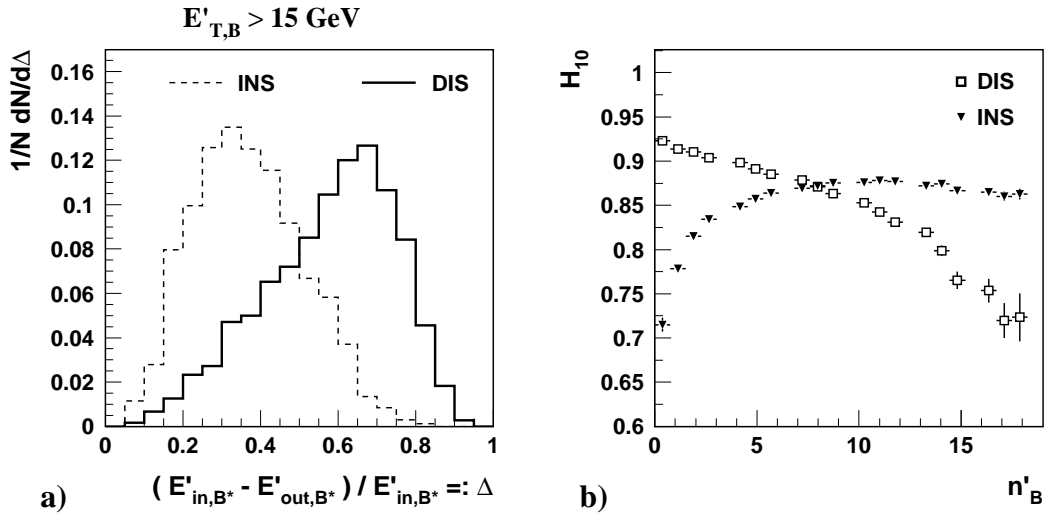,
      width=14cm,bbllx=25pt,bblly=270pt,bburx=570pt,bbury=535,clip=}
    }
   \caption[]
    {a) Shape normalized distributions of the relative 
    $E_{in}$-$E_{out}$ difference (in the instanton band, without jet
    hadrons)
    for normal DIS and I-induced events, while applying
    the additional cut $E'_{T,B} \ge 15$ GeV.
    b) Correlation between the mean first Fox-Wolfram moment $H_{10}$ and the
    number of charged particles $n'_B$ (in the instanton band, without jet
    hadrons).}
   \label{delta-h10}
\end{figure}

This cut-scenario has been
established in a study of what we call ``reference''
(or default) Monte Carlos. For the (default) simulation of normal DIS events
we use the Lund MC generator ARIADNE (\cite{ariadne}, including Pomeron exchange),
which includes the Colour Dipol Model (CDM, \cite{cdm1,cdm2,cdm3})
to describe higher order perturbative QCD radiation. 
Although not perfect, ARIADNE presently gives the best description of the
properties of the hadronic final state at HERA \cite{best,mctun99}.
For the simulation of I-induced events we use
QCDINS~\cite{qcdins1,qcdins3}, which is based on 
the matrix element of the I-subprocess and interfaced by default to 
HERWIG~\cite{herwig} for the simulation of parton showers,
fragmentation and particle decays (see section \ref{qcdins}). 
The shaded band shown in Fig.~\ref{6var} stems from
variations of the Monte Carlo simulations
(see section \ref{depend})\footnote{
A list of {\it all} variations used in creating the shaded band
in Fig.~\ref{6var} is given in Fig.~\ref{var-signal} 
and Fig.~\ref{var-background}.}.\\
All distributions presented here are normalized to the number of events.
Fig.~\ref{6var}a shows the distribution of the $p_T$ of the current
jet, i.e. the jet with the highest $p_T$, for which
$\eta^\ast(jet) \ge \ol{\eta^*}$ holds. The distribution
for I-induced events
is rather flat for $p_T$ values of 1 to $\sim 7$ GeV,
while for normal DIS events the spectrum steeply falls towards higher ${p_T}$.
Fig.~\ref{6var}b shows the distribution of ${Q'}^2_{rec}$, a quantity
which has no direct physical interpretation for normal DIS events,
but which exhibits a rather good separation power (especially in
correlation to $n'_B$, c.\,f. \cite{jgerigk}).
The distribution of the transverse energy and of the
charged particles in the instanton band 
($\ol{\eta^*}\pm 1.1$) is displayed 
in Fig.~\ref{6var}c and in Fig.~\ref{6var}d.
Both distributions reflect the high partonic (and following hadronic)
activity in the I-subprocess.
The observables depicted in Fig.~\ref{6var}e and \ref{6var}f make use of
the shape of I-induced events which is supposed to be more
isotropic than that of normal DIS events. Fig.~\ref{6var}e
gives the distribution of the relative $E_{in}$-$E_{out}$ difference,
a quantity, which measures the $E_T$ weighted $\Phi$ isotropy of an
event (for a detailed description see
\cite{jgerigk,greenshaw}). 
The more homogeneous the transverse energy is distributed
in $\Phi$, the lower the relative 
$E_{in}$-$E_{out}$ difference becomes. Again, we investigate this quantity
in the band\footnote{%As was investigated in \cite{jgerigk}, 
The separation obtained with this quantity can be enhanced by taking
into account all hadrons in a broader range in $\eta^*$ ($\ol{\eta^*}\pm 1.6$). 
This is indicated by the index ``B*''.},
after subtracting the hadrons belonging to the current jet.
Normal DIS events seem to 
exhibit a surprisingly low value of this quantity, implying a rather 
isotropic $\Phi$ distribution of the remaining hadrons.
However, this behaviour strongly varies with the MC model used 
as is illustrated by the shaded band.
This similarity of the distributions for normal DIS
and I-induced events disappears for events with high $E_T$ as
expected and as 
demonstrated in Fig.~\ref{delta-h10}a.
Here, normal DIS
events exhibit the expected jet-like structure with low ($\Phi$) isotropy,
while I-induced events remain as isotropic as before.
%%%%%%%%%%%%%%%%%%%%%%%%%%%%%%%%%%%%%%%%%%%%%%%%%%%%%%%%%%%%%%%%%%%%%%%%%
%%%%%%%%%%%%%%%%%%%%%%%%%%%%%%%%%%%%%%%%%%%%%%%%%%%%%%%%%%%%%%%%%%%%%%%
Finally, in Fig.~\ref{6var}f, we investigate the distributions of
the first Fox-Wolfram moment $H_{10}$ 
\cite{fox} (normalized to the zeroth moment),
an event shape variable which is independent of the axis chosen to 
calculate it.
The Fox-Wolfram moments, in general, seem to perform better than all
other studied event shape variables (like sphericity, thrust etc) 
in separating normal DIS from I-induced events (c.\,f. \cite{jgerigk}).
To understand, why the cut indicated by the arrow in Fig.~\ref{6var}f 
improves the separation power, the correlation between 
$H_{10}$ and $n'_B$ has to be taken into account
(Fig.~\ref{delta-h10}b). For high charged particle 
multiplicities, the mean value of $H_{10}$ of normal DIS events
lies {\it below} that of I-induced events so that cutting at
higher values of $H_{10}$ is reasonable\footnote{The peak of the
$H_{10}$ distribution at $H_{10} \simeq 1$ (Fig.~\ref{6var}f)
is due to diffractive
events (simulated by the Pomeron exchange or the SCI mechanism)
which usually have very low multiplicities ($n\leq 4$).}.\\

The multi-dimensional cut-scenario indicated in Fig.~\ref{6var}
leads to a separation power of $INS_{eff}/DIS_{eff} \simeq 130$,
with the remaining efficiencies
$INS_{eff}\simeq 10\%$ and $DIS_{eff}\simeq 0.08\%$. 
If we extrapolate the prediction for the cross section given in
Eq.~(\ref{WQhera}) to the kinematical range investigated in this report
($x_{\rm Bj}\geq 10^{-4}, Q^2 \geq 5~\mbox{GeV}^2, 0.1<y<0.6$), we
find $\sigma_{HERA}^{(I)} \simeq 215 ~\mbox{pb}$,
and $\sigma _{HERA}^{(I)}\simeq 22 ~\mbox{pb}$ after applying all cuts,
respectively.
For $\mathcal{L} \simeq 30~\mbox{pb}^{-1}$,
one finds for the number of I-induced events $N_{INS}\simeq 670$, while
$N_{DIS}\simeq 1810$ of normal DIS events are expected. 
This implies an $\order{13} \, \sigma$ effect when compared to the 
merely statistical DIS expectation.
%As a result,
%the number of I-induced events remaining would make an effect of
%\order{13} standard deviations, when comparing it to the mere statistical
%fluctuation of the expected number of events.
%
%

%%% Local Variables: 
%%% mode: latex
%%% TeX-master: t
%%% End: 
\section{Dependencies on MC models\label{depend}}
\begin{figure} [ht!]
\centering
\mbox{
   \epsfig{file=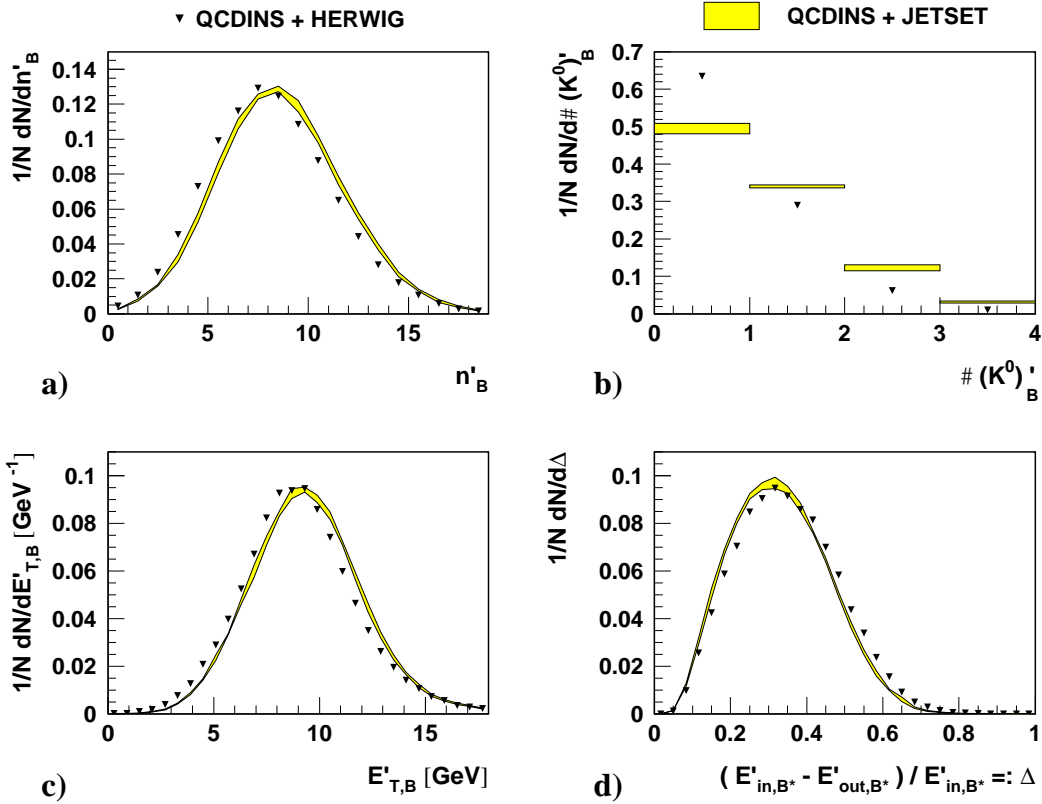,
      width=14cm,bbllx=23pt,bblly=260pt,bburx=555pt,bbury=675,clip=}
    }
   \caption[]
   {Comparison of different models to simulate the hadronisation of
    I-induced events with respect to their effect on a selection of
    distributions.
    The markers represent the default QCDINS implementation
    (QCDINS + HERWIG (tuned) = cluster fragmentation),
    while the shaded band represents QCDINS plus Lund string fragmentation
    (as implemented in JETSET) with different tunings applied (see text).
    }
   \label{her-jset}
\end{figure}

\begin{figure} [ht!]
\centering
\mbox{
   \epsfig{file=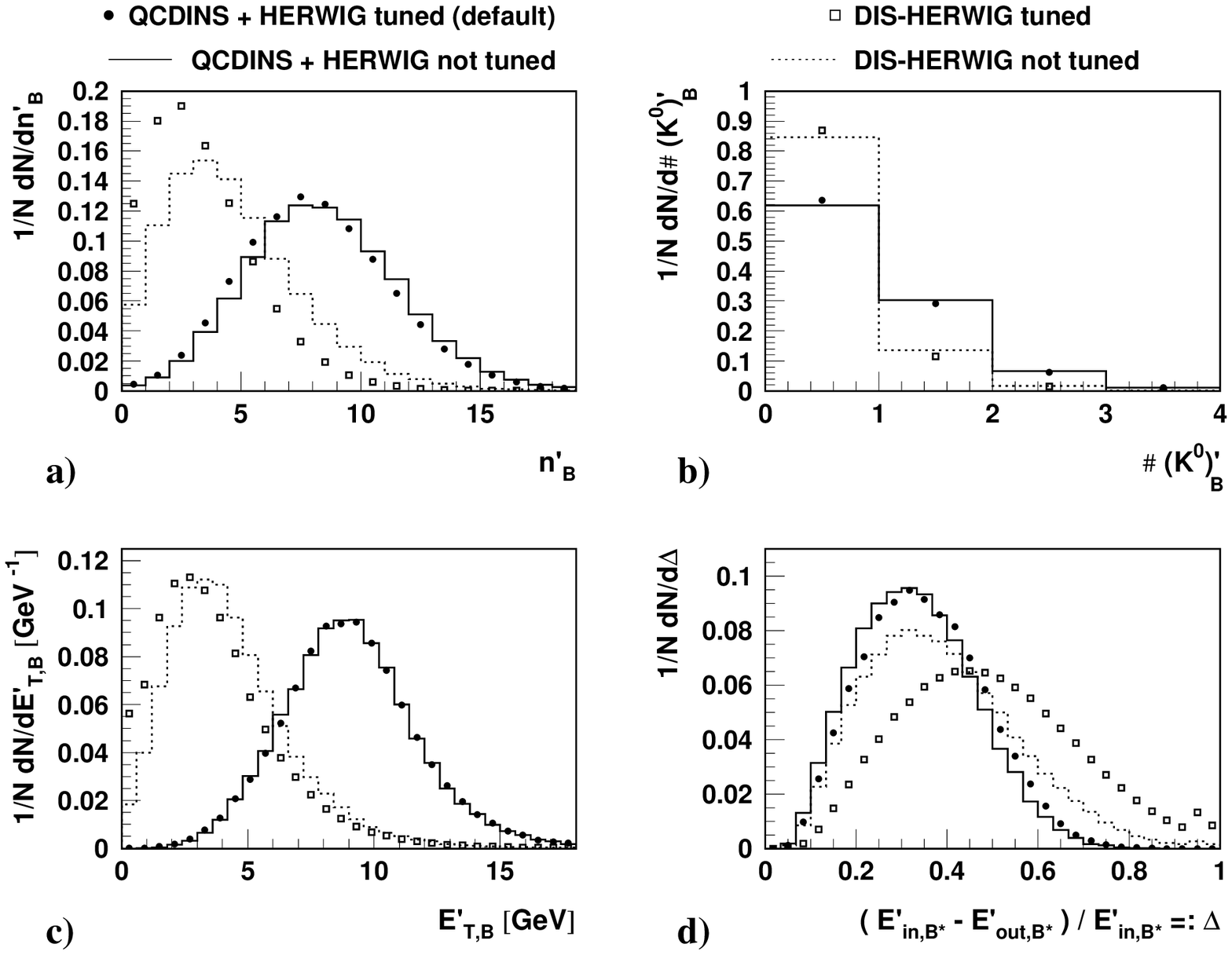,
      width=14cm,bbllx=20pt,bblly=260pt,bburx=570pt,bbury=690,clip=}
    }
   \caption[]
     {Comparison of the influence of HERWIG tuning  
     for normal DIS and I-induced events with respect to its effect on
      a selection of distributions.}
   \label{hertun}
\end{figure}

In this section, we investigate how
%asses the systematic error of the result
%obtained by the cut-scenario presented above, i.e. 
the final-state observables depend on the MC models used for the
simulation of I-induced as well as normal DIS events. 
%To begin with, we investigate the dependence of 
%the characteristics of the hadronic final state of I-induced events 
%on the model used for hadronisation. 
In the default version of QCDINS the cluster 
fragmentation model \cite{cluster} is used.
Given the large number of ${\cal O}(10)$ partons produced in a narrow
$\eta^\ast$-interval by the
I-subprocess, a thorough investigation of hadronization alternatives 
would be very desirable, since so far there is little experience with such
configurations of high parton densities. In the following study
we start with what is available in this context.
For a selection of important distributions, Fig.~\ref{her-jset}
illustrates the effects %evoked by interfacing QCDINS 
%to JETSET \cite{string3}, which 
of using the Lund string model \cite{string3,string1,string2} as implemented
in JETSET \cite{string3} for fragmentation.
%
%%%%%%%%%%%%%%%%%%%%%%%%%%%%%%%%%%%%%%%%%%%%%%%%%%%%%%%%%%%%%%%%%%%%%%%
%%%%%%%%%%%%%%%%%%%%%%%%%%%%%%%%%%%%%%%%%%%%%%%%%%%%%%%%%%%%%%%%%%%%%%%
%
The markers show the default implementation of QCDINS, while the
shaded band reflects variations of QCDINS + JETSET, arising from
different tunings of the JETSET parameters, which are obtained from
studying the hadronic final state in $e^+e^-$ collisions~\cite{eemctun}.
Fig.~\ref{her-jset}a shows that the mean number of charged hadrons
produced in an I-induced event slightly increases when using
string fragmentation. For the number of neutral kaons (cf. 
Fig.~\ref{her-jset}b), however,
%we find a surprisingly large increment in comparison
a significant difference 
to the values obtained in the cluster fragmentation model is found.
The Figs.~\ref{her-jset}c and d basically are a direct result of
the larger number of hadrons produced, i.e. %when using string fragmentation, 
a slightly larger mean value of the transverse
energy and a slightly more ($\Phi$) isotropic distribution of
the particles produced in an I-induced event.
In general, we find that the prediction for the properties of
the hadronic final state of I-induced events depends only weakly
on the choice of the model used to simulate the 
hadronisation.
This statement is supported by studying the effects of the HERWIG tuning,
as suggested in \cite{best}.
Two parameters of the cluster fragmentation model 
influencing the maximum allowed cluster mass (CLMAX) and the
mass distribution of the split clusters (PSPLT)
are changed in the tuned HERWIG version.
In Fig.~\ref{hertun} we find, that this 
modifications strongly influence the distributions for normal DIS 
events\footnote{An exception is the number of neutral kaons, which
seems to be not influenced by this parameter tuning.}.
%%%%%%%%%%%%%%%%%%%%%%%%%%%%%%%%%%%%%%%%%%%%%%%%%%%%%%%%%%%%%%%%%%%%%%%
%%%%%%%%%%%%%%%%%%%%%%%%%%%%%%%%%%%%%%%%%%%%%%%%%%%%%%%%%%%%%%%%%%%%%%%
On the contrary, the distributions of I-induced events only show
rather slight variations\footnote{One exception to this general behaviour is
the distribution of the first Fox-Wolfram moment, presented in
Fig.~\ref{6var}f, which shows rather strong variations when using
JETSET or the untuned HERWIG instead of the tuned HERWIG for the
hadronisation of I-induced events.}.
Besides the influence of the $p_T$ of the current jet, the hadronic final
state of I-induced events seems to be mainly determined by the dynamics of
the I-subprocess, and by the available phase space in $x'$ and $Q'$.

Besides the parameters or models determining the hadronisation, we also
varied the structure function of the proton from CTEQ4L \cite{cteq}
(which we used as default) to MRSH \cite{mrsh} in the simulation of
I-induced events.
This variation results in a small decrease of the total cross section,
but shows almost no visible effect in the distributions
of the hadronic final state observables.
Finally, we varied the scale $\Lambda^{(3)}_{\overline{\rm MS}}$ within 
its experimental
uncertainty as explained in section \ref{signature}. Besides the expected strong 
influence on the cross section of I-induced events, the lower value of
$\Lambda^{(3)}_{\overline{\rm MS}}$ has the effect of slightly enhancing 
the
charged particle multiplicity, while the higher value leads to slightly
lower multiplicities. This effect can be easily explained
as an increase or decrease in the mean gluon multiplicity which is 
proportional to $1/\alpha_s$ and thus increases with decreasing 
$\Lambda^{(3)}_{\overline{\rm MS}}$.

\begin{figure} [ht!]
\centering
\mbox{
   \epsfig{file=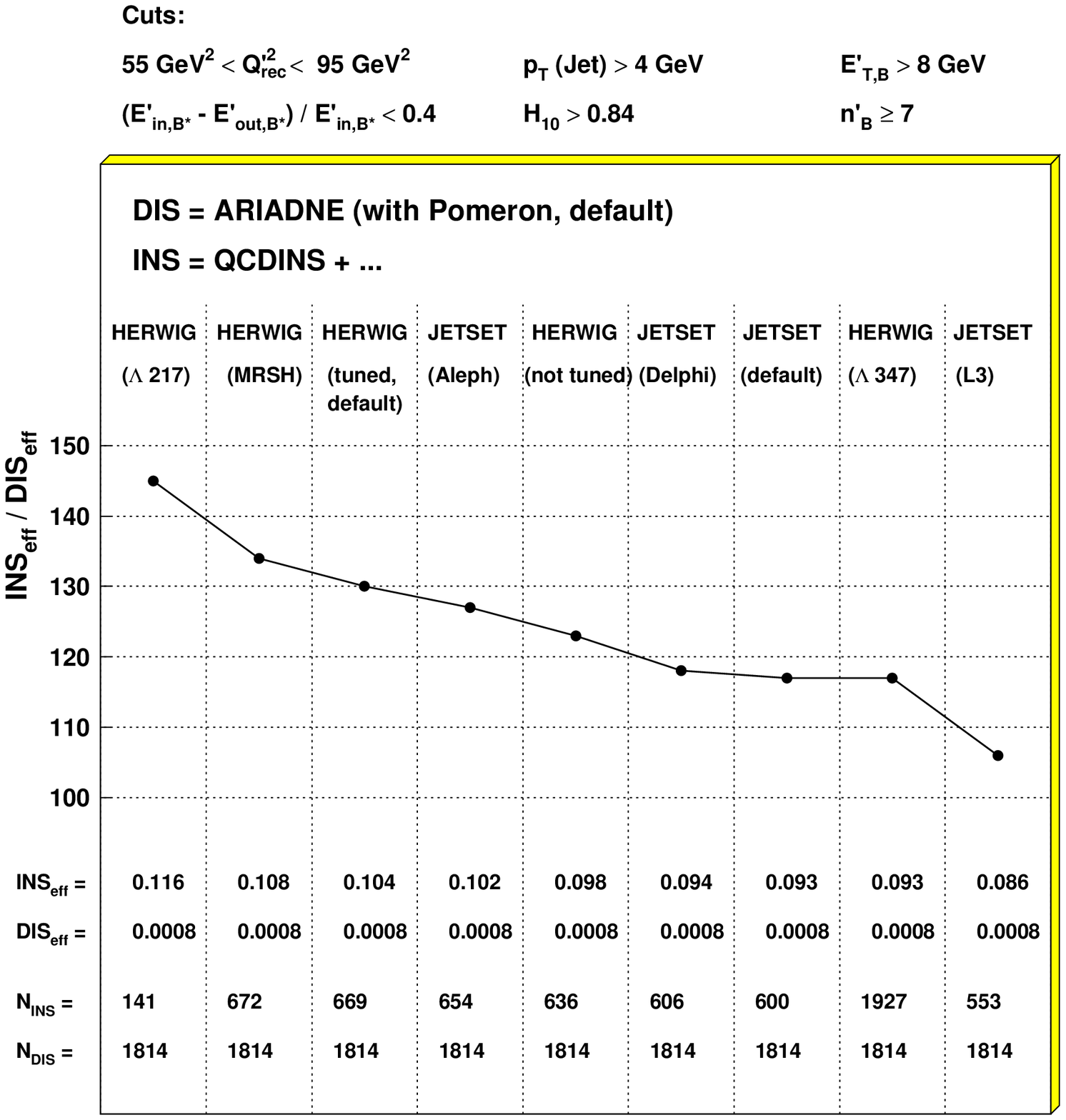,
      width=15cm,bbllx=25pt,bblly=161pt,bburx=530pt,bbury=695,clip=}
    }
   \caption[]
    {Dependence of the separation power ($INS_{eff}/DIS_{eff}$) 
    of a multidimensional cut-scenario on
    the variation of MC models and parameters used to simulate
    I-induced events. The corresponding efficiencies are listed, as are
    the numbers of events remaining (assuming an integrated
    luminosity of $\mathcal{L} \simeq 30~\mbox{pb}^{-1}$).}
   \label{var-signal}
\end{figure}

Fig.~\ref{var-signal} shows the influence of all the variations of the
simulation of I-induced events on the separation power of our
multi-dimensional cut-scenario
(explicitly given on top of the figure).
%%%%%%%%%%%%%%%%%%%%%%%%%%%%%%%%%%%%%%%%%%%%%%%%%%%%%%%%%%%%%%%%%%%%%%%%%
%%%%%%%%%%%%%%%%%%%%%%%%%%%%%%%%%%%%%%%%%%%%%%%%%%%%%%%%%%%%%%%%%%%%%%%
For the simulation of normal DIS events
we use (our default) MC model ARIADNE (with Pomeron exchange) as reference.
As mentioned, the application of all cuts leads to a remaining DIS efficiency
of $8 \cdot 10^{-4}$ or, respectively, a number of events of
$N_{DIS} \simeq 1810$,  when normalized to
$\mathcal{L} \simeq 30~\mbox{pb}^{-1}$. 
The variations used in the modeling of I-induced 
events are sorted according to the separation power. The corresponding
efficiencies range from approx. $9\%$ to $11\%$. After normalization
to the luminosity given above, one finds that the numbers of
I-induced events remaining vary only slightly between $N_{INS}\simeq 550$ and
$N_{INS}\simeq 670$. Because of its strong influence on the cross 
section for
I-induced events, the obvious exception to this is the variation of
the scale $\Lambda^{(3)}_{\overline{\rm MS}}$, which results in the 
worst and the best
prediction for the remaining number of events, namely $N_{INS}\simeq 140$
and  $N_{INS}\simeq 1930$, respectively.

Another important aspect to investigate is the dependence of the separation 
power on variations of the MC models used to
simulate normal DIS events. To this end, three different 
MC generators are used in our study: ARIADNE, LEPTO \cite{lepto} and HERWIG. 
All generators are operated in the ``tuned version'', i.e. with parameter
settings obtained in optimizing the description of the hadronic final
state at HERA \cite{best}.
For the generators ARIADNE and LEPTO, the simulation of diffractive events
(modeled via the Pomeron exchange and the SCI-mechanism, respectively)
has each been turned on and off.
%%%%%%%%%%%%%%%%%%%%%%%%%%%%%%%%%%%%%%%%%%%%%%%%%%%%%%%%%%%%%%%%%%%%%%%%%
\begin{figure} [ht!]
\centering
\mbox{
   \epsfig{file=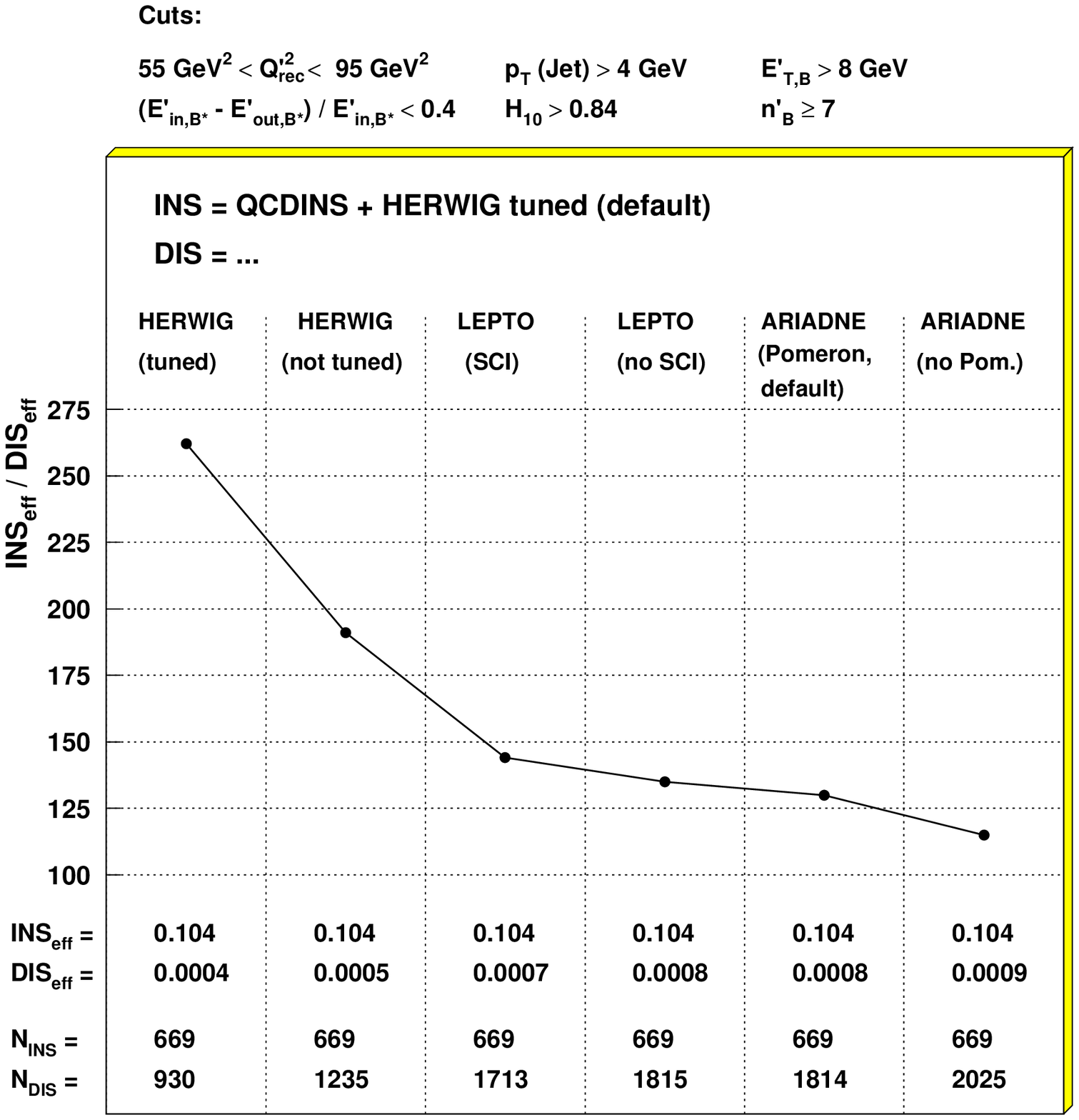,
      width=15cm,bbllx=25pt,bblly=161pt,bburx=530pt,bbury=695,clip=}
    }
   \caption[]
   {Dependence of the separation power ($INS_{eff}/DIS_{eff}$) 
    of a multidimensional cut-scenario on
    the variation of the MC generator used to simulate normal DIS
    events. The corresponding efficiencies are listed, as are
    the numbers of events remaining (assuming an integrated
    luminosity of $\mathcal{L} \simeq 30~\mbox{pb}^{-1}$).}
   \label{var-background}
\end{figure}
%%%%%%%%%%%%%%%%%%%%%%%%%%%%%%%%%%%%%%%%%%%%%%%%%%%%%%%%%%%%%%%%%%%%%%%
Fig.~\ref{var-background} shows the influence of these variations 
on the separation power of our multi-dimensional cut-scenario.
This time, we fix the simulation of I-induced events to be described by
our reference MC model QCDINS + HERWIG (tuned).
The separation powers using LEPTO or ARIADNE turn out
to be rather similar, while HERWIG gives a significantly different result. 
Quite remarkably, the expected difference in separation power,
arising from the usage of different models for hadronisation, is
even more pronounced in the tuned version of HERWIG (which was tuned
according to the same distributions as ARIADNE and LEPTO, c.\,f. \cite{best}).
To demonstrate this effect, the results obtained using the untuned
HERWIG implementation are additionally presented in 
Fig.~\ref{var-background}. %
The obtained efficiencies result (for $\mathcal{L} \simeq 30~\mbox{pb}^{-1}$)
in numbers of remaining normal DIS events ranging from
$N_{DIS}\simeq 930$, at $DIS_{eff}\simeq 4\cdot10^{-4}$
when using HERWIG (tuned), to $N_{DIS}\simeq 2025$, at 
$DIS_{eff}\simeq 9\cdot10^{-4}$, when using ARIADNE (without Pomeron
exchange). For I-induced events
we find the already quoted value of $N_{INS}\simeq 669$.
%%%%%%%%%%%%%%%%%%%%%%%%%%%%%%%%%%%%%%%%%%%%%%%%%%%%%%%%%%%%%%%%%%%%%%%

%\begin{figure} [ht!]
%\centering
%\mbox{
%   \epsfig{file=plots/charm-bottom-def-comp,
%      width=14cm,bbllx=20pt,bblly=260pt,bburx=570pt,bbury=690,clip=}
%    }
%   \caption[]
%     {}
%   \label{charm-bottom}
%\end{figure}
%%%%%%%%%%%%%%%%%%%%%%%%%%%%%%%%%%%%%%%%%%%%%%%%%%%%%%%%%%%%%%%%%%%%%%%

%%% Local Variables: 
%%% mode: latex
%%% TeX-master: t
%%% End: 
\section{Conclusion}
The experimental discovery of instanton-induced processes in DIS
at HERA would be a novel, non-perturbative manifestation of QCD, and
therefore be of basic significance. 
Based on the characteristic hadronic final state of I-induced 
events, we found a set of six observables that exhibit a good
``separation-power'' to normal DIS events. In our best cut-scenario, 
normal DIS events are suppressed by a factor of $8\cdot 10^{-4}$, while
10\% of I-induced events survive.
For $\mathcal{L} \simeq 30~\mbox{pb}^{-1}$ and the nominal value of the theoretical cross section, this results in a predicted number
of $N_{INS}\simeq 670$, for I-induced events, while
$N_{DIS}\simeq 1810$ of normal DIS events are expected.
%Compared to the mere statistical fluctuations of the expected number of
%events, the number of  I-induced events remaining would make an effect of
%\order{13} standard deviations.
%
Within the ``bandwidth'' of variations considered, 
the systematic uncertainties arising from the modeling of I-induced events 
are surprisingly small. In this case, 
the structure of the hadronic final state seems to be mainly
determined by the dynamics of the I-subprocess and by the available phase
space in $x'$ and $Q'$.
On the contrary, the simulation of normal DIS events
is found to be much more sensitive to the variation of model parameters.
Thus, a better understanding of the tails of distributions for normal DIS
events turns out to be quite important. 
\section{Acknowledgments}
We would like to thank % M. Kuhlen for Fig.~\ref{kin-var} and especially 
H. Jung for his help to interface QCDINS with JETSET.
\begin{small}
\bibliographystyle{jetnotit}
\bibliography{literat}
\end{small}
\end{document}

%% file: mydef.tex
\def\shat   {$\hat s $}
\newcommand{\ol}[1]{\mbox{$\overline{#1}$}}
\newcommand{\order}[1]{\mbox{${\cal O}(#1)$}}
\newcommand{\qprimesq}{\mbox{${Q'}^2~$}}
\newcommand{\xprime}{\mbox{$x'~$}}

\newcommand{\Qsq}{\mbox{$Q^2~$}}